\documentclass[twocolumn,twoside,slac_two]{revtex4}
\usepackage{graphicx}
\usepackage{fancyhdr}
\begin{document}

\title{
Hints of the jet composition in gamma-ray bursts
from dissipative photosphere models
}

%

\author{P. Veres}
\affiliation{
The George Washington University, Department of Physics, 725 21st St, NW,
Washington, DC 20052, USA\\
Center for Space Plasma and Aeronomic Research (CSPAR), University of Alabama in Huntsville, Huntsville, AL 35899, USA
}
%




\begin{abstract}
We present a model for gamma-ray bursts where a dissipative photosphere
provides the usual spectral peak around MeV energies accompanied by a
subdominant thermal component.  We treat the initial acceleration of the jet in
a general way, allowing for magnetic field- and baryon dominated outflows.  In
this model, the GeV emission associated with GRBs observed by Fermi LAT, arises
as the interaction of photospheric radiation and the shocked electrons at the
deceleration radius.  Through recently discovered correlations between the
thermal and nonthermal peaks within individual bursts, we are able to infer
whether the jet was Poynting flux or baryon dominated.
\end{abstract}

\maketitle

\thispagestyle{fancy}


\section{Introduction}
Gamma-ray bursts (GRBs) are one of the most extreme phenomena in the universe.
They involve relativistic, jetted emission emanating either from a compact
binary merger or form the core collapse of a massive star. The prompt MeV range
emission component is usually followed by a longer lasting afterglow at lower
energies but sometimes also in the GeV range.

The afterglow can be modeled by shocks propagating in the circumstellar
material \cite{Meszaros+97ag, Sari98}.  It is possible to derive physical
parameters of the late outflow and its surroundings  (e.g. \cite{Pan01}).  

The properties, such as the composition of the relativistic outflow, however,
are difficult find. Dissipative photosphere models \cite{Belob10,Giannios12}
are among the best suited to describe the observed properties of GRB prompt
emission. In these scenarios, energy is dissipated below the photosphere
through some mechanism and released as the optical depth decreases below unity
to produce the prompt MeV range radiation.

In Section 2, we will present the theoretical model for a particular dissipative
photosphere scenario. In Section 3 we will apply this model to the observations of
correlations between thermal and non-thermal components in bright bursts.
Finally we present our conclusions regarding the inferences we can make for the
GRB jet composition.

\section{Model}

\begin{figure*}[t]
\centering
\includegraphics[width=135mm]{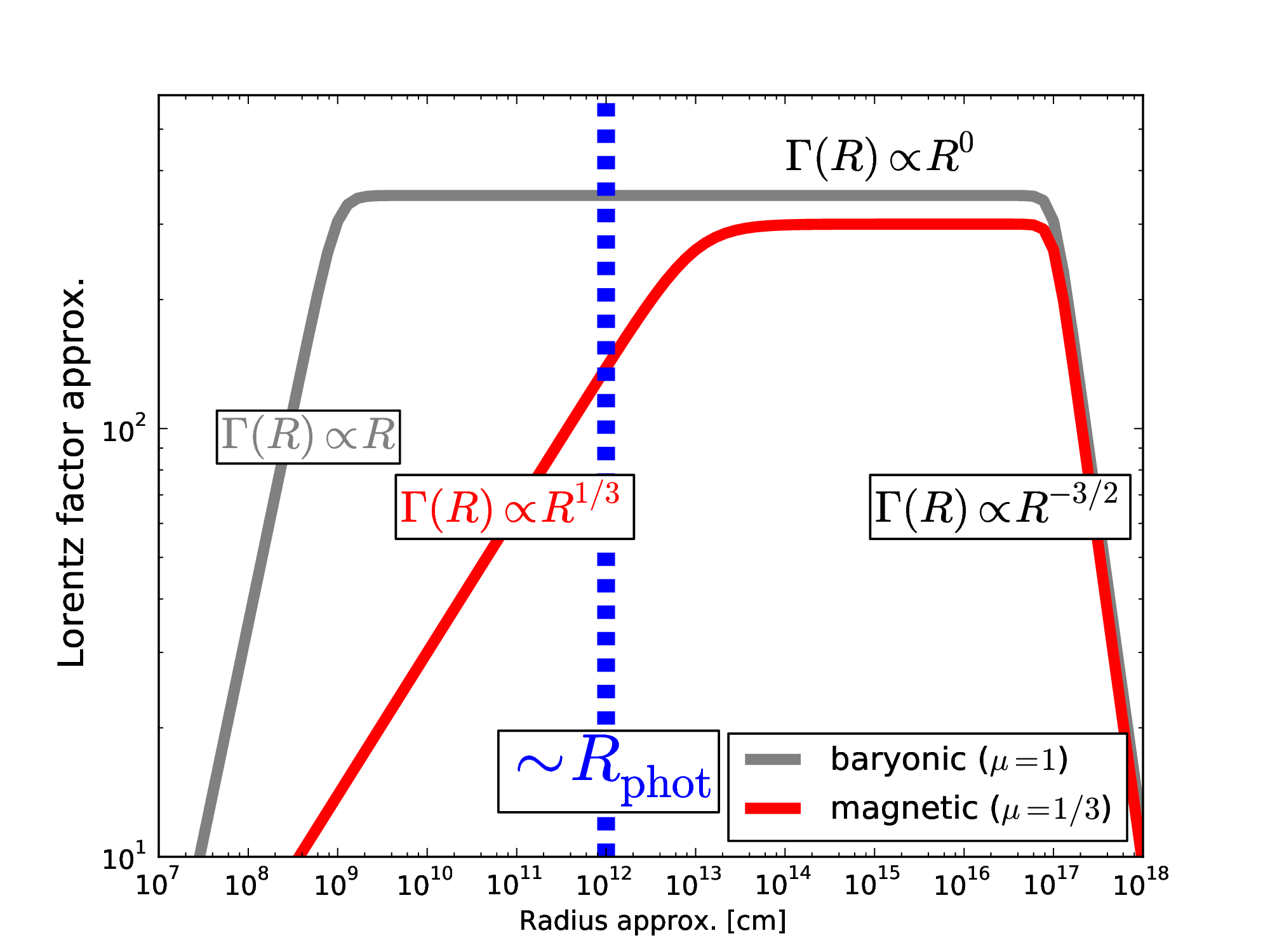}
\caption{Illustration of the Lorentz factor evolution with radius for two
extremal $\mu$
values. Note that the photosphere occurs in the acceleration phase for $\mu=1/3$ and
in the coasting phase for $\mu=1$.} \label{fig1}
\end{figure*}

We model the Lorentz factor of the initial outflow as $\Gamma\propto R^\mu$. If
the energy density of the outflow is dominated by baryons, one expects
$\mu\approx 1$ \cite{Meszaros+93gasdyn}. In case the magnetic fields dominate
the energy budget, one can have an increase as slow as $\mu=1/3$. We introduced
the model in \cite{Veres+12magnetic}, then generalized for arbitrary values of
$\mu$ in \cite{Veres+12fit}.  The acceleration stops at the saturation radius and
the Lorentz factor becomes constant, then decelerates. The start of
deceleration is roughly the start of the afterglow phase (see Figure
\ref{fig1}). 

The main point of such a generalized approach is that the jet can become
optically thin while it is still accelerating. This will happen for $\mu$ close
to $1/3$. The $\mu\lesssim 1$ cases involve photospheres occurring in the
coasting phase.

The photospheric radius, where the outflow becomes optically thin will be at:
\begin{equation}
  \frac{r_{\rm ph}}{r_0}=\left(\frac{ L\sigma_{\rm T}}{8\pi m_{\rm p}
      c^3r_0}\right)\frac{1}{\eta \Gamma_{\rm ph}^2}
\end{equation}
which can be expressed more conveniently by:

\begin{equation}
\frac{r_{\rm ph}}{r_0}=\eta_{\rm T}^{1/\mu}\left\{
\begin{array}{ll}
  (\eta_{\rm T}/\eta)^{1/(1+2\mu)}	& {\rm if~ } \eta>\eta_{\rm T} \\
  (\eta_{\rm T}/\eta)^3		& {\rm if~ } \eta<\eta_{\rm T}
\end{array}
\right. .
\end{equation}
Here $r_0$ is the jet launching radius, $L$ denotes luminosity, $\eta$ is the
coasting Lorentz factor and $\Gamma_{\rm ph}$ is the Lorentz factor at the
photosphere.  $\eta_T$ is a critical Lorentz factor, which discriminates
between the "photosphere in the accelerating phase" and "photosphere in the
coasting phase" cases and can be calculated by equating the saturation radius
to the radius of the photosphere:
\begin{equation}
 \eta_T=\left(\frac{L\sigma_T}{8\pi m_p c^3 r_0}\right)^{\frac{\mu}{1+3\mu}}
\approx \left\{
\begin{array}{lll}
 120~L_{53}^{1/6} r_{0,7}^{-1/6}		&	{\rm if}	& \mu=1/3\\
 1300~L_{53}^{1/4} r_{0,7}^{-1/4}	&	{\rm if}	& \mu=1
\end{array}
\right.
\end{equation}

The main MeV peak will develop close to the photosphere. We model it as
synchrotron radiation from weakly relativistic shocks (involving Lorentz
factors $\Gamma_r\gtrsim 1$).  A sub-dominant thermal component will accompany
the synchrotron peak with characteristic temperature in the range of 1 keV-100
keV in accordance with observations.

The synchrotron peak will have the following dependence on the intrinsic physical parameters:
\begin{eqnarray}\label{epeak}
E_{\rm peak} \propto \left\{
\begin{array}{ll}
  L^{\frac{3\mu-1}{4\mu+2}}~ \eta^{-\frac{3\mu-1}{4\mu+2}}~
  r_0^{\frac{-5\mu}{4\mu+2}}~ \Gamma_r^3		&	{\rm if~ }
  \eta>\eta_T	\\
  L^{-1/2}~ \eta^{3}~ \Gamma_r^3		&	{\rm if~ }
  \eta<\eta_T,
\end{array}
\right.
\label{eq:peak}
\end{eqnarray}

while the thermal component will have the following dependence:

\begin{eqnarray}\label{temp}
T_{\rm obs} \propto \left\{
\begin{array}{ll}
  L^{\frac{14\mu-5}{12(2\mu+1)}}~ \eta^{\frac{2-2\mu}{6\mu+3}}~
  r_0^{-\frac{10\mu-1}{6(2\mu+1)}} 		&	{\rm if~ }
  \eta>\eta_T\\
  L^{-5/12}~ \eta^{8/3}~ r_0^{1/6}  	&	{\rm if~ }
  \eta<\eta_T.
\end{array}
\right.
\end{eqnarray}
The above scenario is able to fit bright LAT detected bursts
\cite{Veres+12fit}, but cannot discriminate between the magnetic and baryonic
cases.

\section{Application of the model to observations}
\begin{figure}[t]
\centering
\includegraphics[width=\columnwidth]{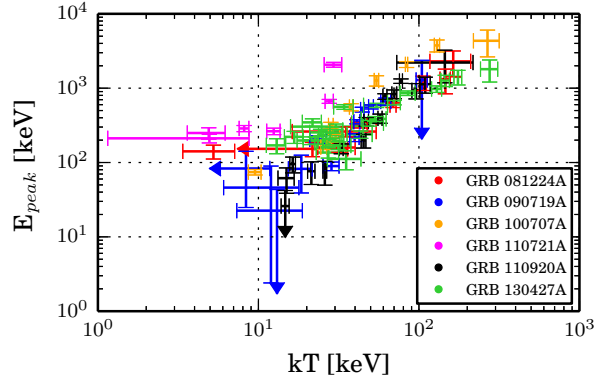}
\caption{The correlation between the synchrotron peak energy and the thermal
peak for six bursts with time-resolved spectra (Figure from \cite{Burgess:2014b}). } \label{fig2}
\end{figure}

\cite{Burgess:2014} developed a model that includes a synchrotron and a thermal component. This model can 
successfully fit the spectra of GRBs. By fitting this model to  bright
GRBs, \cite{Burgess:2014b} found a correlation between the peak energy of the
synchrotron and the peak of the thermal component, $E_{\rm peak}\propto
T^{\alpha}$. Every burst in the sample has characteristic $\alpha$ index (see
Figure \ref{fig2} and Table \ref{tab:comp}).

Using the model described in the previous section, we can link the theoretically derived peak energy with
the temperature of the thermal component (Equations \ref{epeak} and \ref{temp}).
We can carry out this exercise either through the luminosity (L) or the coasting Lorentz
factor($\eta$) and get similar expressions. For the "photosphere in the
acceleration phase" we have: 
\begin{equation}
E_{\rm peak} \propto  T^{\frac{6(3\mu-1)}{14\mu-5}},
\end{equation}
while if the "photosphere is in the coasting phase", we have: 
\begin{equation}
E_{\rm peak}
\propto  T^{1.2}.
\end{equation}
Now, it is a simple task to identify the observed $\alpha$ indices with the
indices in the above  equations.

We compile the results in table \ref{tab:comp} for our sample of six bursts.
For bursts with higher values of $\alpha$ we derive a $\mu$, which points
towards a magnetic origin in the cases of GRBs 090719A, 100707A and 110920A
respectively. In the remaining cases, the $\alpha$ index does not depend on
$\mu$. Nonetheless, the values of $\alpha$ for GRBs 081224A, 110721A and
130427A are very close to the theoretically derived 1.2. This points to the
"photosphere in coasting phase" scenario, which in turn is easiest to realize
in the baryon dominated case (because for reasonable parameters the critical
Lorentz factor is $\eta_T \gtrsim 1000$).

\begin{table}[t]
\begin{center}
\caption{Results for determining jet composition}
\begin{tabular}{|l|c|c|c|}
\hline
   GRB Name   &           $\alpha$  & Jet Type & $\mu$   \\
\hline
  GRB 081224A  	&	$1.01   \pm	0.14	 $ &{baryonic}   &   $-$         \\
  GRB 090719A   &	$2.33   \pm	0.27	 $ &{magnetic}   & 0.39$\pm$0.01 \\
  GRB 100707A  	&	$1.77	\pm	0.07	 $ &{magnetic}   & 0.42$\pm$0.01  \\
  GRB 110721A	&	$1.24	\pm	0.11	 $ &{baryonic}   &     $-$        \\
  GRB 110920A   &	$1.97	\pm	0.11	 $ &{magnetic}   & 0.4$\pm$0.01  \\
  GRB 130427A 	&	$1.02	\pm	0.05	 $ &{baryonic}  	&     $-$      \\

\hline
\end{tabular}
\label{tab:comp}
\end{center}
\end{table}

\section{Conclusion} 
We presented a model where the initial acceleration of the gamma-ray burst
outflow is written as a function of a parameter ($\mu$). This parameter
characterizes the composition of the outflow.  By deriving a relation between
the synchrotron peak energy and the temperature of the thermal component from
the model, we were able to explain the observed relation between the two
quantities and infer the composition of the gamma-ray burst outflow. 

Half of the sample appears magnetically dominated while the other baryon
dominated. Thus, there are no obvious trends among bright bursts regarding
their composition.

\bigskip 
\begin{acknowledgments}
The author thanks P. M\'esz\'aros, K. Dhuga, W. Briscoe, J. M. Burgess, B.-B.
Zhang for their help and collaboration.  This work was partially supported by
the OTKA NN 111016 grant. The author further thanks the kind support of the
Physics Department and the Astro Group at the George Washington University.
\end{acknowledgments}

\bigskip 

\begin{thebibliography}{9}   


\bibitem{Belob10}
Beloborodov, A.~M. 2010, MNRAS, 407, 1033

\bibitem{Burgess:2014}
Burgess, J.~M. {et~al.} 2014, ApJ, 784, 17

\bibitem{Burgess:2014b}
Burgess, J.~M. {et~al.} 2014, ApJ, 784, 43

\bibitem{Giannios12}
Giannios, D. 2012, MNRAS, 422, 3092

\bibitem{Meszaros+93gasdyn}
M\'esz\'aros, P., Laguna, P., \& Rees, M.~J. 1993, ApJ, 415, 181

\bibitem{Meszaros+97ag}
{M{\'e}sz{\'a}ros}, P. \& Rees, M.~J., 1997, ApJ, 476, 232 

\bibitem{Pan01}
Panaitescu, A, \& Kumar, P., ApJL, 560, 49 

\bibitem{Sari98}
Sari, R., Piran, T. \& Narayan, R., 1998, ApJL, 497, 17

\bibitem{Veres+12fit}
Veres, P., Zhang, B.~B., \& M\'esz\'aros, P. 2013, ApJ, 764, 94

\bibitem{Veres+12magnetic}
{Veres}, P. and {M{\'e}sz{\'a}ros}, P. 2012, ApJ, 755, 12 



\end{thebibliography}

\end{document}